\documentclass[amsmath,amssymb,prb,preprint]{revtex4}

\usepackage{graphicx}
\begin{document}

\title{Dimensional crossover of thermal conductance in nanowires }

\author{Jian Wang}
\author{Jian-Sheng Wang}

\affiliation{Center for Computational Science and Engineering and
Department of Physics, National University of Singapore, Singapore
117542, Republic of Singapore}

\date{April 5, 2007}

\begin{abstract}
Dimensional dependence of thermal conductance at low temperatures in
nanowires is studied using the nonequilibrium Green's function
(NEGF) method. Our calculation shows a smooth dimensional crossover
of thermal conductance in nanowire from one-dimensional to
three-dimensional behavior with the increase of diameters. The
results are consistent with the experimental findings that the
temperature dependence of thermal conductance at low temperature for
diameters from tens to hundreds nanometers will be close to Debye
law. The calculation also suggests that universal thermal
conductance is only observable in nanowires with small diameters. We
also find that the interfacial thermal conductance across Si and Ge
nanowire is much lower than the corresponding value in bulk
materials.

\end{abstract}


\maketitle

Thermal properties of semiconductor nanowires  have attracted
significant attention in recent years with the continuous scaling
down of feature sizes in microelectronic devices and
circuits.\cite{deyuli, superlattice, measure, schwab} Semiconductor
nanowires promise applications in future generation electronic and
optoelectronic devices. The reduced dimension effects on thermal
transport in nanowires becomes  important  both for device
reliability and for intrinsic physics. For one-dimensional quantum
atomic chains\cite{schwab}, thermal conductance is proportional to
the temperature $T$ at low temperatures with the quantized universal
coefficient $ \pi^2 k_B^2/3h$.  For three-dimensional bulk
materials, it is well known that thermal conductance depends on
temperature as Debye $ T^3$ law at low temperatures. With the
dimension falling between one and three dimensions,  the behavior of
thermal transport in quasi-one-dimensional nanowires will be an
interesting problem to be investigated. Recent experimental results
of thermal conductance \cite{deyuli} in Si nanowires with the
diameter $\rm 22\,$nm
 exhibited a deviation from Debye law. The
temperature dependence of thermal conductance \cite{measure} in Si
nanowires with a cross section of $\rm 130\!\times\!200 \, nm^2$ was
shown to behave as $T^3$ above ${\rm 1.2 \, K}$.  It is, therefore,
important to systematically explore the dimensional effects on
thermal conductance in nanowires.

In this letter, we model thermal transport in nanowires  using the
 nonequilibrium Green's function method
(NEGF)\cite{haug,PRB-green,yamamoto,our-green}. We find a
one-dimensional to three-dimensional  transition of thermal
conductance at low temperatures in nanowires. Interfacial thermal
conductance across Si and Ge nanowires is found substantially
smaller in comparison with the value across bulk Si and Ge epitaxial
interface.

 We consider the Si semiconductor nanowires as an example. Unlike the three-dimensional bulk
materials, there is no translational invariance in the transverse
direction in a nanowire. We choose a conventional supercell that
includes all the atoms in the transverse directions. Nanowire
structures are first optimized using Tersoff
potential\cite{Tersoff}. Force constants for each atom are obtained
from the equilibrium position under small displacements. We have
verified that the force constants from Tersoff potential  reproduce
reasonably well the phonon dispersion of bulk silicon
material\cite{giannozzi}.  Thermal transport in nanowires is
calculated along the $[100]$ direction.

Phonon transport in nanowires is treated using the nonequilibrium
Green's function  formalism, as described in
Ref.~\onlinecite{haug,PRB-green,yamamoto,our-green}.  Thermal
current expression for the lead, for example the left lead, is given
by the formula as
\begin{equation}
I = - \frac{1}{2\pi}\!\! \int_{-\infty}^{+\infty}\!\!\!\!\!\!\!\!
d\omega\, \hbar \omega \, {\rm Tr}\Bigl( G^r[\omega]
\Sigma^{<}_L[\omega] + G^{<}[\omega] \Sigma^a_L[\omega] \Bigr),
\label{heat-current}
\end{equation}
where $G^r[\omega]$ and $G^{<}[\omega]$ are the retarded and the
lesser Green's function for the scattering region, respectively. The
 subscript $L$ denotes the left lead.  The lesser self-energy
$\Sigma^{<}_L$ and the advanced self-energy $\Sigma^a_L$ account for
the coupling of the scattering region with the left lead. Similar
expressions can be written down for thermal current on the right
lead. The retarded Green's function $G^r$ is obtained from the
solution of the Dyson equation, as
\begin{equation}
\label{dysoneq} G^r[\omega]=\Bigl ( (\omega+i0^{+})^2{\bf
I}-K_c-\Sigma _R^r -\Sigma _L^r -\Sigma _n^r \Bigr ) ^{-1},
\end{equation}
where $K_c$ is the dynamic matrix for the central scattering region.
Here $\Sigma _L^r , \Sigma _R^r $ and $  \Sigma _n^r
 $ are the retarded self-energies due to the coupling with the left/right
lead and from the nonlinear phonon-phonon interaction, respectively.
The retarded self-energy for the left/right lead $\Sigma _{\alpha}^r
 ,(\alpha=L,R)$, is calculated through the relation $\Sigma_{\alpha}^r=V_{\alpha}g_\alpha^{r}
 V^{\dagger}_{\alpha}, \alpha=L,R$. Here $g_\alpha^{r}$ is the surface
 Green's function which can be calculated through a recursive
 iteration  method\cite{our-green}. The matrix $V_{\alpha}$ is the coupling matrix
 between the semi-infinite lead and the central region. The
 nonlinear self-energy $\Sigma _n^r$  can be computed through the
 expansion of Feynman diagrams or the mean-field theory\cite{our-green}. Here we concentrate on the
 dimensional dependence of thermal transport in nanowires at low
 temperatures. We will ignore the nonlinear interactions. This approximation
  is reasonable because the phonon boundary instead of the phonon-phonon scattering dominates  thermal transport at moderation high temperatures in nanowires\cite{deyuli}.
  If the nonlinear self-energy is not  considered, Eq.~(\ref{heat-current}) can be further reduced to the
 Landauer formula\cite{PRB-green,yamamoto,our-green}. The thermal conductance $G=\frac{1}{S_0}\partial I/\partial T$ ,where $S_0$ is the cross-section of the nanowire,
 in terms of
 Landauer formula is given as
 \begin{equation}
 \label{equ:caroli}
 G=\frac{1}{2\pi\,S_0} \int_{0}^{\infty}\!\!\!\!d\omega\,\hbar \omega
 \,\mathcal{T}[\omega]\frac{\partial f}{\partial T},
  \end{equation}
  where the Caroli transmission is $\mathcal{T}[\omega]={\rm Tr}(G^r\Gamma_LG^a\Gamma_R)$ and
  $\Gamma_\alpha=i(\Sigma^r_\alpha-\Sigma^a_\alpha),\alpha=L,R$.

\begin{figure}[ht]
\includegraphics[width=0.70\columnwidth]{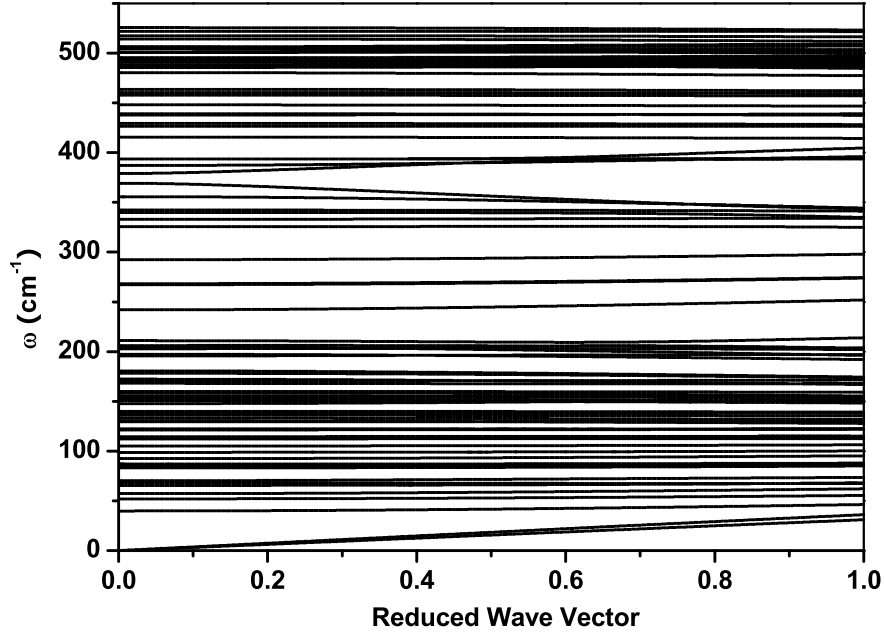}
\caption{\label{fig:phonondispersion} Phonon dispersions of Si
nanowire with the diameter $d=1.54 \,{\rm nm}$. The force constants
are calculated from Tersoff potential. The wave vectors are in terms
of the reduced wave number of the first Brillouin zone. }
\end{figure}

  The phonon dispersion relation for Si nanowire with the diameter $d=1.54 \, {\rm nm}$ is
  illustrated in Fig.~\ref{fig:phonondispersion}.  It can be seen from Fig.~\ref{fig:phonondispersion} that there are
  three acoustic branches: one longitudinal branch and two degenerately
  transverse branches. The maximum frequency for acoustic branches in Si nanowire is $36\, {\rm
  cm^{-1}}$, which is in contrast with the value of about $340 \, {\rm cm^{-1}}$
  for Si bulk\cite{giannozzi}. It can be seen for Fig.~\ref{fig:phonondispersion} that many optical branches with small group velocities
  emerge.

\begin{figure}[ht]
\includegraphics[width=0.600\columnwidth]{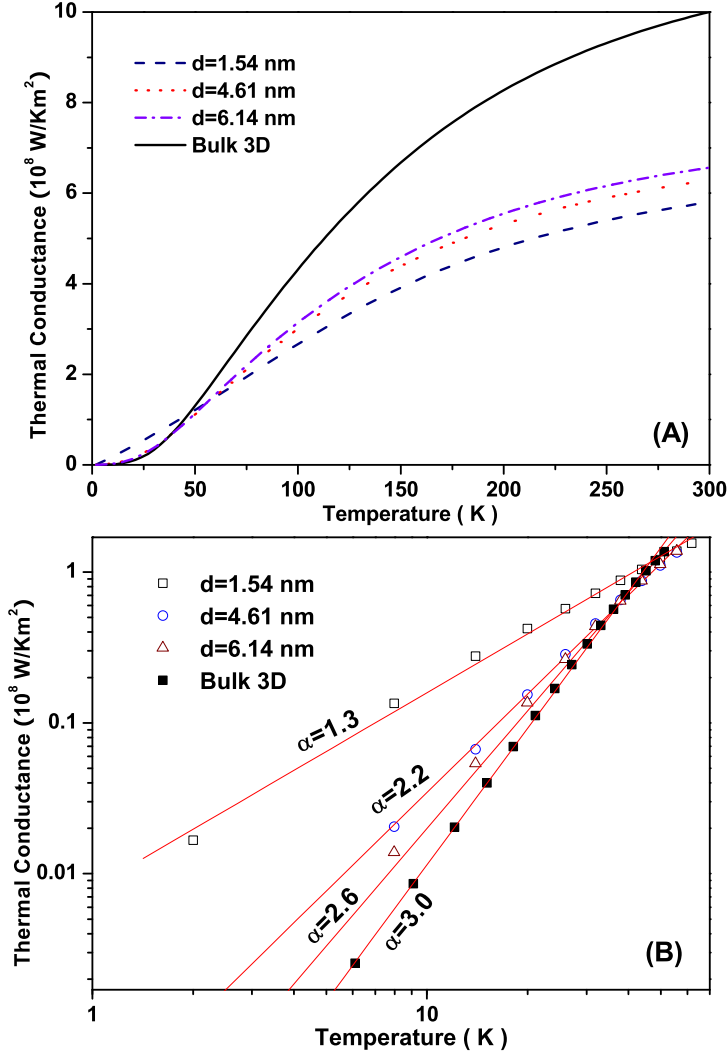}
\caption{\label{fig:thermalconductance}  \textbf{(A)} Thermal
conductance for Si nanowires with different diameters and for Si
bulk. \textbf{(B)} The log-log plot of thermal conductance at low
temperatures.    }
\end{figure}

  Thermal conductance calculated from NEGF for Si nanowires with the
  diameters ranging from $1.54 \, {\rm nm}$ to $ 6.14 \, {\rm nm}$ is shown in
  Fig.~\ref{fig:thermalconductance}.   For comparison, thermal conductance for bulk silicon
  calculated with the force constants derived from Tersoff
  potential is also plotted in Fig.~\ref{fig:thermalconductance}.   When temperature $T>60K$,
  thermal conductance increases with the diameters of Si nanowires,
  but it is still below the value of bulk Si material. It can be explained by the fact that the optical branches which dominate thermal conductance in Si nanowires have smaller group velocities.
   This kind of diameter-dependent behavior is also consistent with the experimental results in Ref.~\onlinecite{deyuli}.
  The most significant feature in Fig.~\ref{fig:thermalconductance} is the temperature dependence of thermal conductance with the
  increase of the diameter of nanowires at low temperatures. The temperature dependence of the thermal conductance $G\propto T^{\alpha}$  below $ {\rm 60} $K is plotted on a log-log scale in
  Fig.~\ref{fig:thermalconductance}(B). It can be seen from Fig.~\ref{fig:thermalconductance}(B) that the
exponent $\alpha$ changes from $\alpha = 1.3$ to $\alpha = 2.6$ with
the increase of diameters. This dimensional crossover of thermal
conductance from one-dimensional $T$ behavior to three-dimensional
Debye $T^3$ law is clearly seen in
Fig.~\ref{fig:thermalconductance}(B). It can be concluded from this
figure that universal thermal conductance can only  be observed in
nanowires with very small diameters.

\begin{figure}[ht]
\includegraphics[width=0.700\columnwidth]{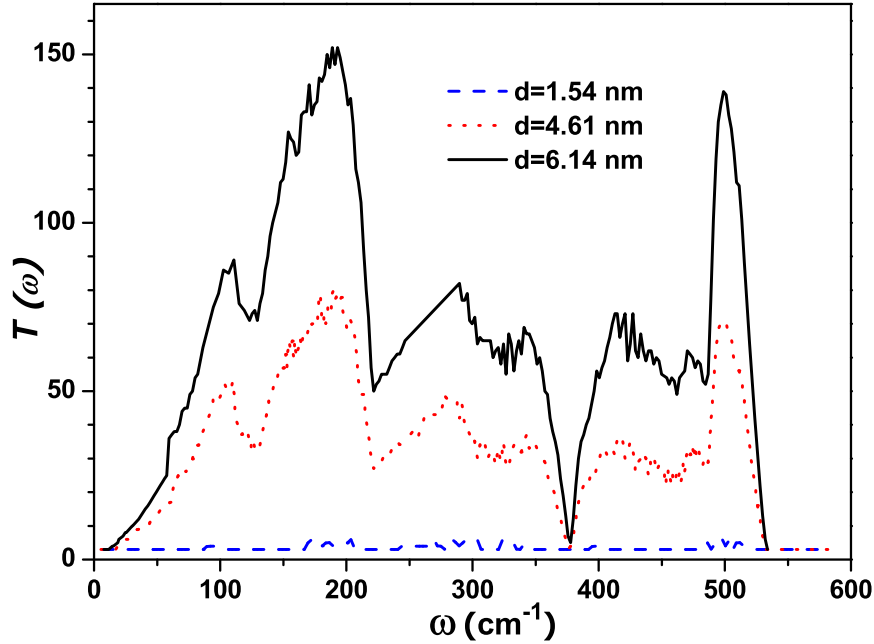}
\caption{\label{fig:dos} The dependence of transmission function
$\mathcal{T}$ on frequencies for Si nanowires with different
diameters. }
\end{figure}

To understand the above dimensional crossover behavior, we plot the
transmission $\mathcal{T}[\omega]$ dependence on frequency $\omega$
for Si nanowires with different diameters in Fig.~\ref{fig:dos}.
We can assume that  the transmission $\mathcal{T}$ has a dependence
on frequency with the relation
$\mathcal{T}\!\!\!\propto\!\omega^\beta$ in the range of low
frequencies. For one dimension, the transmission equals $1$ so that
$\beta=0$. For three-dimensional bulk, ballistic thermal conductance
$G_{bulk}$ at low temperature can be written as
$G_{bulk}=\frac{1}{(2\pi)^3}\Sigma_s \int d^3{\bf q} \, \hbar
\omega_s \, v_s^z\frac{\partial f}{\partial T}$, where $s$ denotes
the different polarized branch. Compared with
Eq.~(\ref{equ:caroli}), the transmission
$\mathcal{T}_{bulk}[\omega]$ for bulk material is given as
 \begin{equation}
 \label{3dtran}
\mathcal{T}_{bulk}[\omega]= \frac{1}{(2\pi)^2}\Sigma_s \!\!\int
\!d^3\!{\bf q} \,\delta(\omega-\omega_s({\bf q}))v_s^z.
\end{equation} Note that $\mathcal{T}_{bulk}[\omega]$ in Eq.(\ref{3dtran}) is equivalent to $\mathcal{T}[\omega]/S_0$ in Eq.(\ref{equ:caroli}).
In the low-frequency range, only the acoustic branches need to be
considered for three-dimensional bulk materials. If we use the Debye
model, then all three branches of the spectrum have the linear
dispersion relation as $\omega_s=c_sq$. Substituting this relation
into Eq.(\ref{3dtran}),  we find that the transmission for bulk
material is given as
$\mathcal{T}_{bulk}[\omega]=\Sigma_s(\frac{\omega}{c_s})^2/4\pi\propto
\omega^2$.  Thus the transmission function at low frequency for
three dimensional-bulk  depends quadratically on the frequency, {\rm
i.e.} $\beta=2$. It is a straightforward conclusion from
Eq.(\ref{equ:caroli}) that thermal conductance $G$ at low
temperature will behave as $T^{\beta+1}$, \textit{i.e.}
$\alpha=\beta+1$, if the transmission function depends on frequency
as $T^\beta$.   It can be seen from Fig.~\ref{fig:dos} that the
exponent $\beta$ for the transmission at low frequency increases
with the diameters of nanowires. This explains why the temperature
dependence of thermal conductance at low temperature increases from
$\alpha=1.3$ to $\alpha=2.6$ with the diameters from $d=1.54 \,{\rm
nm}$ to $d=6.14 \, {\rm nm}$.

\begin{figure}[ht]
\includegraphics[width=0.70\columnwidth]{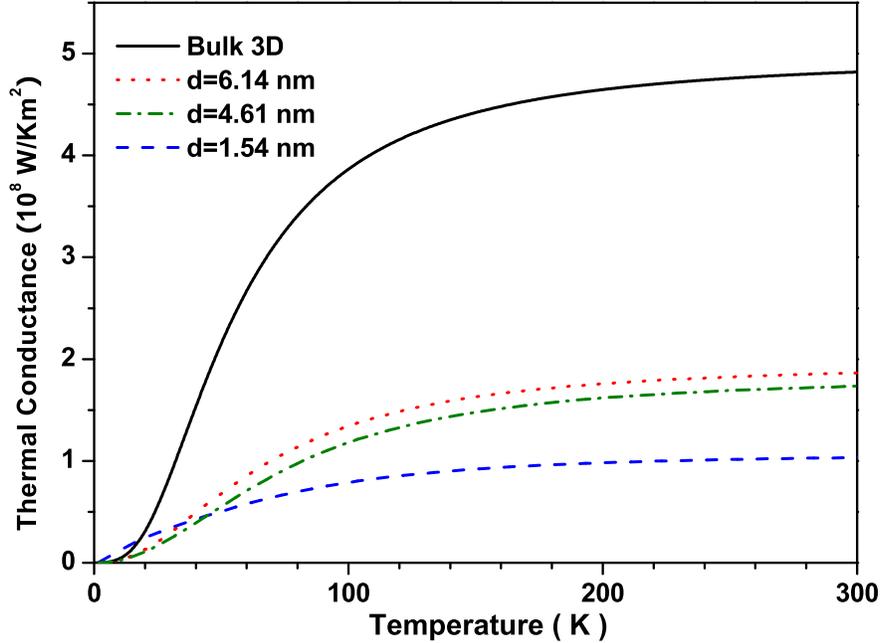}
\caption{\label{fig:mfpfigure} Temperature dependence of interfacial
thermal conductance for Si-Ge nanowires with different diameters
${\rm d}=1.54,4.61,6.14 \,{\rm nm}$ and for the epitaxial interface
between bulk Si and Ge (the solid line). }
\end{figure}

Furthermore we investigate interfacial thermal conductance in
nanowires using NEGF.  Thermal conductance in the Si and Ge
superlattice nanowires\cite{superlattice} was found very small in
comparison with the value of bulk materials. Our aim is to
understand the role played by the Si-Ge interface in nanowires. The
force constants for Si and Ge used in the calculation are also
derived from Tersoff potential after the optimization of structures.
The thermal conductance across silicon and germanium nanowires with
different diameters is plotted in Fig.~\ref{fig:mfpfigure}.  For
comparison, epitaxially interfacial thermal conductance between bulk
silicon and germanium is calculated using the mode-matching lattice
dynamic method\cite{dayang, jwang}. It can be seen from
Fig.\ref{fig:mfpfigure} that interfacial thermal conductance across
Si and Ge nanowires shows a similar dimensional crossover of
temperature dependence like that of pure Si nanowires at temperature
below ${\rm 60 K}$. When the temperature is above  ${\rm 60 K}$,
interfacial thermal conductance increase with the diameters of
nanowires. Thermal conductance at temperature ${\rm T=200K}$ between
Si and Ge nanowires with the diameter ${\rm d=6.14nm}$ is
$1.7\times10^8 {\rm W/Km^2}$, which is about one third of thermal
conductance $4.6\times10^8 {\rm W/Km^2}$ at the same temperature
between bulk Si and Ge epitaxial interface. In contrast with thermal
conductance $5.6\times10^8 {\rm W/Km^2}$ in pure Si nanowire with
the same diameter, thermal conductance across Si-Ge nanowires is
about one fourth of it. This substantially decreased thermal
conductance across Si and Ge nanowires may result from the reduced
group velocity for optical branches in nanowires.

In summary, thermal conductance in nanowires is calculated using
NEGF. Our calculation shows a clear dimensional crossover for
temperature dependence of thermal conductance in nanowires at low
temperatures. We conclude that thermal conductance at low
temperature in most experimental nanowires with diameters $d\!>\!10
\,{\rm nm}$ will behave close to $T^3$. At moderately high
temperature, thermal conductance will increase with the nanowire
diameter and tends to the upper limit  for the corresponding bulk
material. The small value of interfacial thermal conductance across
Si-Ge nanowires is one of the reasons that account for substantial
reduction of thermal conduction in superlattice nanowires. Our
present calculation holds at low, or moderate, temperature when
phonon-phonon scattering does not play a dominant role in thermal
conductance in low-dimensional materials. When temperature is high
enough, nonlinear phonon-phonon scattering should be included.

We thank  Jingtao L\"u and  Nan Zeng for careful reading of the
manuscript. This work is supported in part by a Faculty Research
Grant of National University of Singapore.


\end{document}